\documentclass[aps,prl,amsmath,amssymb,nofootinbib, preprintnumbers]{revtex4}

\usepackage[caption=false]{subfig} 
\usepackage{hyperref} 
\usepackage{graphicx,xcolor}
\usepackage{epstopdf} 
\usepackage{xcolor} 

\def\pa{{\partial}}
\newcommand{\be}{\begin{equation}}
\newcommand{\ee}{\end{equation}}
\newcommand{\ba}{\begin{eqnarray}}
\newcommand{\ea}{\end{eqnarray}}
\newcommand{\nn}{\nonumber}

\newcommand{\h}[1]{\hat{#1}} 
\newcommand{\bb}[1]{\bar{#1}}

\newcommand{\hv}{\hbar v_F}

\begin{document} 
\title{Electronic properties of curved few-layers graphene: a geometrical approach}

\author{Marco Cariglia}
\affiliation{Departamento de F\'isica, Universidade Federal de Ouro Preto, 35400-000 Ouro Preto MG, Brazil} 
 
\author{Roberto Giamb\`o}
\affiliation{School of Science and Technology, Mathematics Division, University of Camerino, 62032 - Camerino, Italy} 
 
\author{Andrea Perali}
\affiliation{School of Pharmacy, Physics Unit, University of Camerino, 62032 - Camerino, Italy}


\begin{abstract} 
We show the presence of non-relativistic L\'evy-Leblond fermions in flat three- and four-layers graphene with AB stacking, extending the results obtained in \cite{Curvatronics2017} for bilayer graphene. When the layer is curved we obtain a set of equations for Galilean fermions that are a variation of those of L\'evy-Leblond with a well defined combination of pseudospin, and that admit L\'evy-Leblond spinors as solutions in an approriate limit. 
The local energy of such Galilean fermions is sensitive to the intrinsic curvature of the surface. 
We discuss the relationship between two-dimensional pseudospin, labelling layer degrees of freedom, and the different energy bands. 
For L\'evy-Leblond fermions an interpretation is given in terms of massless fermions in an effective 4D spacetime, and in this case the pseudospin is related to four dimensional chirality. A non-zero energy band gap between conduction and valence electronic bands is obtained for  surfaces with positive curvature.
\\ 

\vspace{0\baselineskip} \noindent\textbf{Keywords}: 
Few-layers Graphene; L\'evy-Leblond equations; non-relativistic fermions, Eisenhart lift; curved systems. 
\\

\end{abstract}

\maketitle

\section{Introduction}

Graphene is a two-dimensional (2D) material which enables
the realization of different heterostructures and electronic devices with novel
quantum properties \cite{Geim2013,Perali2013,Perali2014}.
The electronic and transport properties of graphene
in a multilayer stacking can be modified and tuned by changing the number
of layers of graphene in the system. Remarkably, the low energy electronic
band structure of multilayered graphene evolves from Dirac cones in
monolayer graphene with massless Weyl excitations,
to parabolic bands with massive excitations in the case of bilayer graphene,
to more complex band structures and peaked density of states when a
few-layer graphene system is realized with different stacking orders \cite{McCann2006,Min2008,Koshino2009,Zhang2010}.
Few-layer graphene can be fabricated from graphite
by mechanical exfoliation \cite{Ferrari2006,Zhang2005}
and by chemical techniques \cite{Berger2004,Shih2011,Mahanandia2014}
controlling the stacking order. Experimental characterizations of
electronic and transport properties of trilayer graphene have been reported
in Ref. \cite{Craciun2009,Bao2011,Mak2010}.
An electric field perpendicular to the graphene layers has been
shown experimentally to open a band gap in the single-particle electronic
energy spectrum in bilayer \cite{Zhang2009} and
trilayer \cite{Lui2011,Zou2013} graphene, being this effect of
crucial importance in realizing semiconducting behavior for electronics.
Theoretical predictions of a similar band gap opening in the energy
spectrum of few-layer graphene have been reported \cite{Avetisyan2010}. 
The electronic band structure of graphene and many other 2D systems can be also tailored by strain of their lattice in different directions, inducing several possible quantum effects. 
Strain plays a key role in iron-based superconductors \cite{Poccia2010}, in cuprate 
superconductors \cite{Agrestini2003}, and in superconducting diborides \cite{Agrestini2001}.
In cuprate superconductors the displacement of the Cu ions along the transversal
direction induces strain modulations and bending in the copper-oxide planes in the form of lattice stripes leading to amplification effects of the superconducting parameters \cite{Bianconi96a,Bianconi96b}.
Curved 2D systems are a source of different kind of compressive and tensile strains, with
a wide range of strain percentages of the lattice depending on the topology and on the curvature
radius of the surface of the curved systems. Therefore we expect that curvature will be
another possible way to engineer the electronic band structure of 2D systems, as we will
discuss in this work. 
Recently the low energy electronic properties of bilayer graphene have been studied in the context of a geometrical approach,
 using the Lévy-Leblond equation \cite{LevyLeblond1967} for massive particles \cite{Curvatronics2017} in a curved space.
Although the study of graphene through relativistic approaches is a quite new field of research, a significant amount of literature already exists, in the study of deformed monolayer graphene employing the relativistic 2D Dirac equation in a curved metric \cite{Sitenko2007,Cortijo2007,deJuan2007,Cortijo2007b,Vozmediano2010,Arias2015,Amorim2016}. 
This connection between graphene and relativity has been explored also in other contexts, for instance for the evolution of the free electron current density in graphene
with defects \cite{sepheri2017,Capozziello2018}.
On the other hand, the scientific challenge to connect Dirac theory of the spinning electron with gravitation
dates back to the seminal works of Hermann Weyl \cite{Weyl1929}. Interestingly,
Weyl fermions with zero mass and definitive chirality have not been identified
in high-energy physics, but they have been clearly observed in
novel topological materials in condensed matter systems \cite{Wan2011}.
The role of two-dimensionality in generating Weyl states in systems
with parabolic electronic bands has been demonstrated in Ref.\cite{Doria2017},
while the influence of zero helicity states at the interfaces of
layered heterostructures, leading to local magnetization and mass anisotropy,
has been discussed in Ref.\cite{Rodrigues2017}.
The above described example demonstrates that solid state systems with
geometrical constraints and specific topological properties can work as
model systems for physical phenomena with very different energy scales,
as high-energy physics and gravitation in a curved D-dimensional space-time.
The mapping proposed in Ref.\cite{Curvatronics2017}
allowed to study the evolution of the band structure,
both conduction and valence bands and their band-gap, as a function of
geometrical curvature. Positive (spherical-like) curvatures have a local effect of opening a band gap
in the spectrum, while negative (hyperbolic-like) curvatures tend to close
the band gap. The band-gap energy is predicted to be tunable and proportional
to the curvature radius of the deformation. This paves the way to
``Curvatronics'' with bilayer graphene, an interesting possibility for
applications to control in a static way the electronic properties
of layered systems with the curvature.
In this work we extend the analysis of Ref.\cite{Curvatronics2017}, analysing massive and massless fermionic excitations in the more complicated case of few-layer flat graphene systems, while for the case of the curved bilayer we refine the treatment, proposing a set of equations that take into account the exact combination of pseudospin states that defines physical states, i.e. eigenstates of the energy. Similar considerations apply to multi-layer systems and will be studied separately.

The work is organised as follows. In sec. \ref{sec:GalileanFermions} we discuss Galilean fermions in few layers graphene: we begin in \ref{sec:two_layers} recalling the results obtained in \cite{Curvatronics2017} for bilayer graphene, and then in \ref{sec:three_layers} and \ref{sec:four_layers} we show that Galilean fermions are present in the case of three and four layers as well. We obtain the exact spinor solutions for all energies, and study them in the proximity of the \textit{Galilei points}, where the electronic dispersions acquire their extreme values: the results is given by 2D spinors that satisfy the Galilei invariant L\'evy-Leblond equation, and we show where these fermions are localised. Then in sec. \ref{sec:3} we discuss various aspects of the L\'evy-Leblond equation, both in the flat and curved case. We recall its Galilean invariance and discuss its two independent solutions, which correspond to states with different pseudo-spin.  It has been discussed in \cite{Curvatronics2017} how solutions of the L\'evy-Leblond equation in 2D can be lifted to solutions of the massless Dirac equation in 4D. We show here how states with a definite pseudospin lift to states with definite chirality, with a perfect match of degrees of freedom: the states corresponding to positive and negative isospin are mapped into the two solutions of the massless 4D Dirac equation, one with left and one with right chirality. Then we discuss in detail the case of the curved bilayer. We present a set of covariant equations that describe the generalization of the flat massive fermions, show how the equations predict that energy eigenstates are given by a well defined mixture of pseudospin states and infer that the scalar curvature of the surface alters the local energy density.

\section{Galilean fermions in few layers graphene\label{sec:GalileanFermions}}

\subsection{Bilayer graphene\label{sec:two_layers}} 
In this section we review the results obtained in \cite{Curvatronics2017} for bilayer graphene with AB Bernal stacking, presenting them in a manner that is suitable to generalisation to a higher number of layers. 
 
Bilayer graphene with AB stacking presents one set of metallic bands $E_1^{(\pm)}$ , and another of isolating bands $E_2^{(\pm)}$. 
We define a quantity with the dimensions of mass 
\be 
m_0 = \frac{\gamma}{v_F^2} \, , 
\ee 
where $\gamma$ is the interlayer hopping parameter, $\gamma \sim 0.4$eV, and $v_F \sim 10^6$ms${}^{-1}$ is the Fermi velocity in graphene. 
The energy bands can be described in terms of $m_0$ and of the effective mass $m = \pm \frac{m_0}{2}$ for bilayer graphene, where the $\pm$ factor denotes either positive or negative bands: 
\be \label{eq:energy_free1}
E_i^{(\pm)} = \pm \left[ (-1)^{i} |m| v_F^2 + \sqrt{ m^2 v_F^4 + |\hv\kappa|^2} \right] \, , i = 1,2 ,.  
\ee  
It is useful to define a dimensionless factor $\alpha$ by 
\be \label{eq:alpha}
\alpha |m| := m_0  
\ee 
such that, in the bilayer case, $\alpha = 2$. 
The $\bf{k} \cdot \bf{p}$ model that describes these bands is defined in terms of a spinor $\lambda$ with 4 components. The Hamiltonian is given by 
\be \label{eq:H}
H_{\kappa}^{(AB)} = \left( \begin{array}{cccc} 
0 & \hv\bb{\kappa} & 0 & 0 \\ 
\hv \kappa & 0 & \gamma & 0 \\ 
0 & \gamma & 0 & \hv \bb{\kappa} \\ 
0 & 0 & \hv \kappa & 0 
\end{array} \right) \, , 
\ee 
where $\kappa = \tau k_x + i k_y$ is the wave number of the excitation, with $\tau = \pm 1$ denoting the two inequivalent Fermi points\footnote{Notice that, with respect to \cite{Curvatronics2017}, we have used a different although equivalent basis for the Hamiltonian, exchanging the vectors $1\leftrightarrow 2$ and $3\leftrightarrow 4$. This will result, for instance, in slightly different expressions for eqn. \eqref{eq:sol_bilayer_spinors1}.}. These are the points in the Brillouin zone where the conduction and valence bands touch, and for graphene they are at the level of the Fermi energy. The $\bf{k} \cdot \bf{p}$ model can be obtained in three steps: 1) assuming that both the action of the full Hamiltonian, and the overlap between wavefunctions, is restricted to nearest neighbour contributions (the tight-binding model restricted to the nearest-neighbor hoppings gives a satisfactory description of the electronic structure of bilayer graphene at low energies, see e.g. \cite{Jung2014}), 2) expanding the momentum close to the Fermi points, and 3) assuming that the most relevant modes for the expansion of the full wavefunction $\Psi$ of  a single electron are given by $p_z$ orbitals, where $z$ is the direction transverse to the bilayer, as these are the orbitals that contribute the most to conduction. In formulas 
\be 
\psi \sim \lambda_1 \left| p_z ; A1 \right. \rangle + \lambda_2 \left| p_z ; B1 \right. \rangle + \lambda_3 \left| p_z ; A2 \right. \rangle + \lambda_4 \left| p_z ; B2 \right. \rangle \, , 
\ee 
where $A, B$ correspond to inequivalent types of carbon atoms in the same plane, and the index $1, 2$ corresponds to different layers. Given the form \eqref{eq:H} of the Hamiltonian, we can recognise that $B1$ and $A2$ are the two atoms that are directly aligned along the transversal direction. Solving the eigenvalue equation $H \lambda = E \lambda$ we find that there are no non-trivial eigenspinors with $\lambda_1 = 0$ when $|\kappa|\neq 0$. The component $\lambda_1$ can then be set to $1$ as the eigenvalue equation is homogeneous. The full solution follows from linear algebra 
\ba \label{eq:sol_bilayer_spinors1}
\lambda_2 &=& \frac{E}{\hv\bar{\kappa}} \, , \quad \lambda_3 = \sigma \frac{E}{\hv \bb{\kappa}} \, , \quad  
\lambda_4 =  \sigma \frac{\kappa}{\bb{\kappa}} \, ,  
\ea 
where $\sigma=\pm 1$ is a factor labelling different bands that appear in the consistency condition satisfied by the energy 
\be 
E^2 - |\hv \kappa |^2 - \sigma \gamma E = 0 \, . 
\ee

In \cite{Curvatronics2017} we expanded the solution around the Fermi points in the dimensionless parameter $\epsilon= \frac{\hv|\kappa|}{\gamma}$.  For the metallic $E_1^{(\pm)}$ bands one obtains 
\ba 
\lambda_2 &=& \frac{\hbar \kappa}{2 m v_F} \label{eq:bilayer_metallic_lambda2} +O\left(\epsilon^2\right)  \, , \\ 
\lambda_3 &=& \mp \frac{2}{\alpha} \frac{\hbar \kappa}{2 m v_F} + O \left( \epsilon^2\right) \, , \label{eq:bilayer_metallic_lambda3} \\ 
\lambda_4 &=& \mp \frac{2}{\alpha} \frac{\kappa}{\bb{\kappa}} \, . \label{eq:bilayer_metallic_lambda4}
\ea 
For simplicity we set $\tau = +1$ and focus on one of the Fermi points. To begin, it is useful noticing that  for all values of the constant $A$ the following spinors 
\ba 
\label{LL_solution1_E1_tau+}
 \chi_1(t)  &=& 
 e^{- i \frac{\hbar k^2 t}{2m}}  \left( A   \lambda_3 ,  \lambda_2 \right)^T \, , \\ 
 \label{LL_solution2_E1_tau+}
\chi_2(t)  &=& 
 e^{- i  \frac{\hbar k^2 t}{2m}}  \left(  \lambda_1 ,A \lambda_4   \right)^T \, , 
\ea 
satisfy the L\'evy-Leblond equations \cite{LevyLeblond1967} 
\ba \label{eq:LevyLeblond1}
&&  i \hbar \, \pa_t \chi_2 + i \hbar v_F D \chi_1 = 0 \, , \\ 
&& D \chi_2 - i \frac{2 m v_F}{\hbar} \chi_1 = 0 \, , \label{eq:LevyLeblond2}
\ea 
where $D = i \sigma^j k_j$ is the $2$--dimensional Dirac operator in phase space and the $\sigma^j$ are the Pauli matrices in the standard basis. As we will discuss in sec. \ref{sec:3}, these are coupled, first order equations that are invariant under the Galilei group of non-relativistic mechanics and describe non-relativistic spin $\frac{1}{2}$ fermions. In particular, the spinors with $\lambda_3 = 0 = \lambda_4$ correspond to a solution with definite pseudospin, and those with $\lambda_1 = 0 = \lambda_2$ to an independent solution with opposite pseudospin. We notice from the explicit form of the solution (\ref{eq:bilayer_metallic_lambda2}-\ref{eq:bilayer_metallic_lambda4}) that, if in the solution we exchange at the same time the energy $E_1^+$ with $E_1^-$, keeping $\kappa$ unchanged, then the components $\lambda_1$ and $\lambda_3$ of the solutions are unchanged, while $\lambda_2$ and $\lambda_4$ change sign. Thus, by taking a linear combination of such eigenstates of energies $E_1^+$ and $E_1^-$ it is possible to obtain `checkered' states where either $\lambda_2 = 0 = \lambda_4$, i.e. only sites of type $A$ are excited, or viceversa states with $\lambda_1 = 0 = \lambda_3$, i.e. only sites of type $B$ are excited. Conversely, taking a linear combination of the latter it is possible to build eigenstates of the energy. We also notice that, as explicit electronic excitations are defined by their Fourier coefficients for each value of the momentum $\kappa$, these are free parameters in our analysis, which concerns only the energy spectrum. This is why we are able to set $\lambda_1 = 1$ at $|\kappa | \neq  0$. In sec.\ref{sec:3} we will see how these results are generalized in the case of curved sheets. We will obtain equations for Galilean fermions that are a variation of the L\'evy-Leblond equation, which involve a well defined combination of pseudospin. We will see that, at least in the axially symmetric case, it is still true that linear combinations of states with opposite energies give checkered states.

\begin{figure}%
         \centering
         \subfloat[$\lambda_2=\lambda_3=0$]{\label{fig:overlap-b}\includegraphics[width=0.45\textwidth]{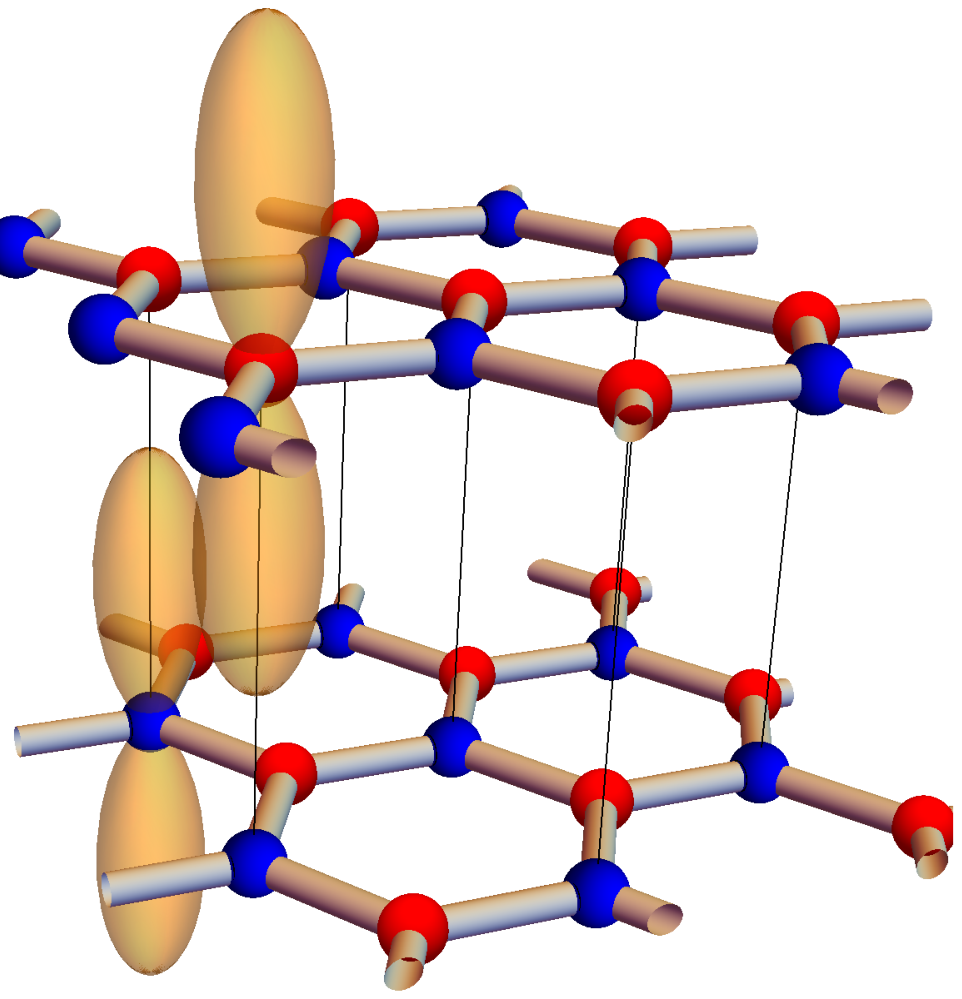}}\qquad
         \subfloat[$\lambda_1=\lambda_4=0$]{\label{fig:overlap-a}
         \includegraphics[width=0.45\textwidth]{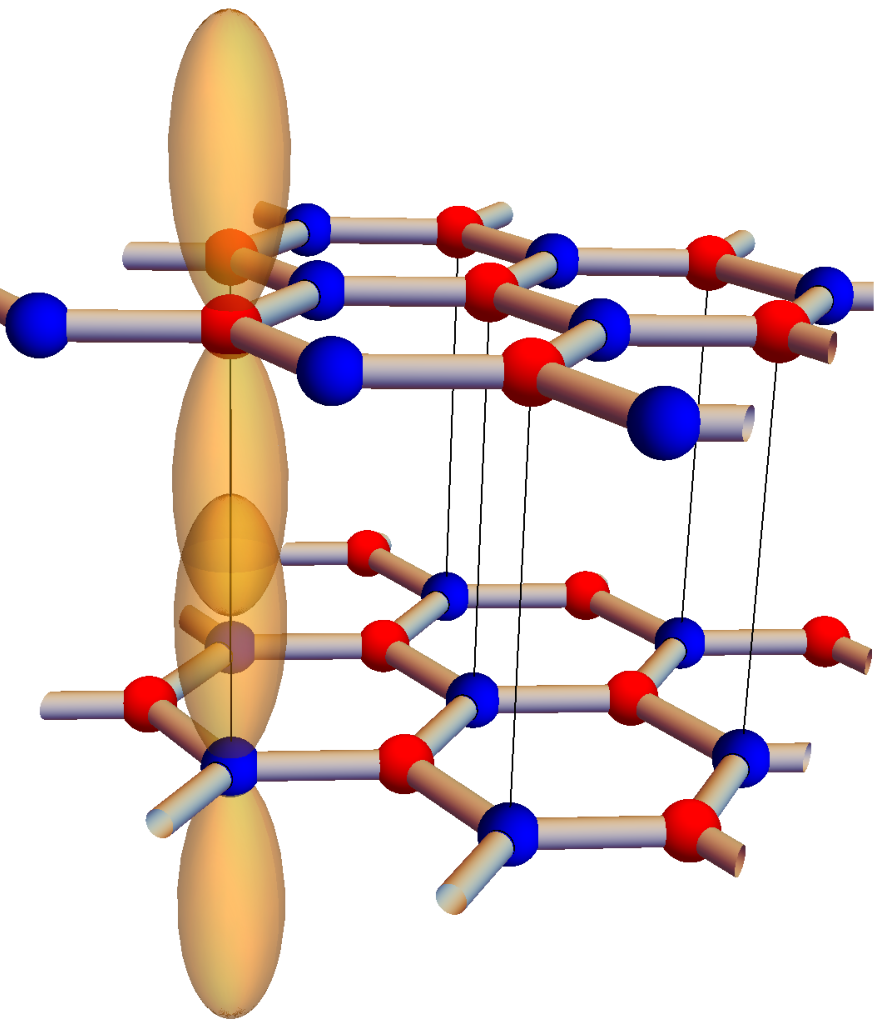}} 
         \caption{Schematic representation of the overlap of the $p_z$ orbitals of bilayer graphene at the Galilei points. The two panels correspond to different choices of the eigenvectors.\ }
         \label{fig:overlap}
       \end{figure}

 In analogy with the fact that for monolayer graphene the Fermi points are called Dirac points because of the relativistic emergent symmetry, for the bilayer we will use the wording \textit{Galilei points}. At the Galilei point, i.e. for $\kappa = 0$, the solutions have $\lambda_2 = 0$, $\lambda_3 = 0$. As shown in Figure \ref{fig:overlap-b}, this corresponds physically to activating only the $p_z$ orbitals of the $A_1$ and $B_2$ atoms: these are not directly aligned along the $z$ direction, thus giving a small overlap of the electron orbitals.  
 
Conversely, when we perform a similar analysis for the insulating bands $E_2^{(\pm)}$ at the same $\tau = +1$ Galilei point we find  
\ba 
\lambda_2 &=&  \frac{2 m v_F}{\hbar \bb{\kappa}} + O \left( \epsilon^2\right) \label{eq:bilayer_insulating_lambda2}  \, , \\ 
\lambda_3 &=& \pm \frac{2}{\alpha} \frac{2 m v_F}{\hbar \bb{\kappa}}+O \left( \epsilon^2\right) \label{eq:bilayer_insulating_lambda3} \, , \\ 
\lambda_4 &=& \pm \frac{2}{\alpha} \frac{\kappa}{\bb{\kappa}} \label{eq:bilayer_insulating_lambda4} \, .  
\ea
and 
\ba 
\label{LL_solution1_E2_tau+}
 \chi_1(t)  &=& 
 e^{- i \frac{\hbar k^2 t}{2m}}  \left(  \lambda_1 , A \lambda_4    \right)^T \, , \\ 
 \label{LL_solution2_E2_tau+}
\chi_2(t)  &=& 
e^{- i \frac{\hbar k^2 t}{2m}}  \left(  A \lambda_3 , \lambda_2   \right)^T \, .  
\ea 
This time, at the Galilei point $\lambda_1 = 0$, $\lambda_4 = 0$ and only the orbitals of the $B_1$, $A_1$ atoms are activated, as shown in Figure \ref{fig:overlap-a}. For the $E_2^{(+)}$ bands the wavefunctions of the two orbitals are in phase, maximising the overlap energy, and for the $E_2^{(-)}$ are out of phase, thus minimising it.

\subsection{Three-layer graphene\label{sec:three_layers}}
\begin{figure}%
         \centering
         \subfloat[$3$--layer]{\label{fig:flat3}\includegraphics[width=0.45\textwidth]{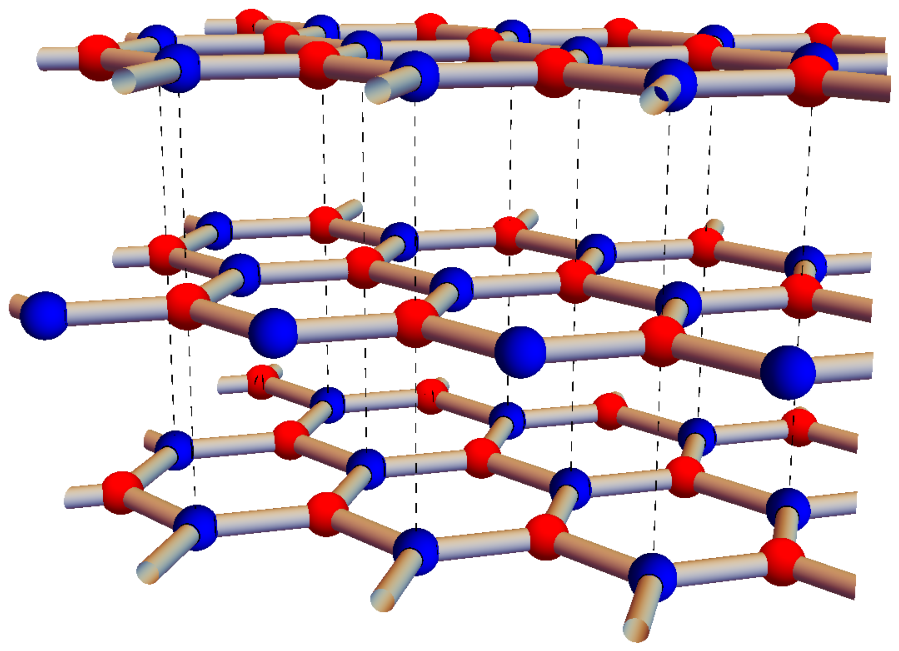}}\qquad
         \subfloat[$4$--layer]{\label{fig:flat4}
         \includegraphics[width=0.45\textwidth]{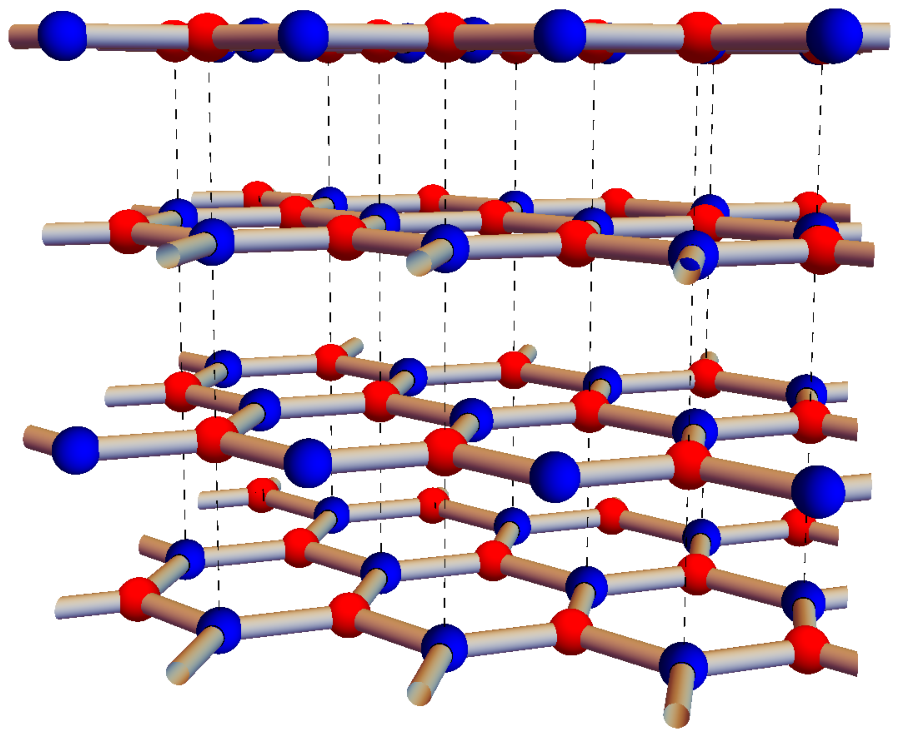}} 
         \caption{(a). Schematic representation of the three-layer graphene lattice with the ABA stacking. (b) Four-layer graphene lattice with the ABAB stacking.}
         \label{fig:flat}
  \end{figure}

Here we extend our approach to the three-layer graphene with the ABA stacking
of the layers, which is represented in Figure \ref{fig:flat3}. 

The effective low energy Hamiltonian for ABA stacked graphene around the K point can be found in \cite{KatnelsonEtAl_2011_Trilayer}: 
\be \label{eq:ABA_Hamiltonian}
H_\kappa^{(ABA)} = \left( 
\begin{array}{cccccc} 
	0 & \hbar v_F \bb{\kappa} & 0 & 0 & 0 & 0 \\ 
	\hbar v_F \kappa & 0 & \gamma & 0 & 0 & 0 \\ 
	0 & \gamma & 0 & \hbar v_F \bb{\kappa} & 0 & \gamma \\ 
	0 & 0 & \hbar v_F \kappa & 0 & 0 & 0 \\ 
	0 & 0 & 0 & 0 & 0 & \hbar v_F \bb{\kappa} \\ 
	0 & 0 & \gamma & 0 & \hbar v_F \kappa & 0 
\end{array} 
\right) \, . 
\ee 
Here we see that the atoms that are directly aligned along the transverse direction are $B_1$ and $A_2$, as before, and $A_2$ and $B_3$. Analysing the band structure one finds two touching bands linear in momentum 
\be 
E_{ABA,0} = \pm\hbar  v_F |\kappa| \, , 
\ee 
and a set of metallic and insulating bands given by 
\be \label{eq:EABA}
\hspace{-0.35cm} E_{ABA,i}^{(\pm)} = \pm \left[ \gamma^2 + \hbar^2 v_F^2 |\kappa|^2 + (-1)^i \gamma \sqrt{\gamma^2 + 2  \hbar^2 v_F^2 |\kappa|^2} \right]^{\frac{1}{2}} \hspace{-0.25cm},  \quad i = 1, 2 \, .  
\ee 
We begin by analising the metallic and insulating bands: we perform an analysis near the Fermi points in order to describe the detailed structure of the spinors. This will allow us to build spinors that satisfy the L\'evy-Leblond equations. We expand the energy bands in the dimensionless parameter $\epsilon = \frac{\hbar v_F |\kappa|}{\gamma}$, finding for the two metallic bands 
\be \label{eq:ABA_metallic_low_energy}
E_{ABA,1}^{(\pm)} \sim  \frac{\hbar^2  |\kappa|^2}{2m} + O\left( \epsilon^4 \right) \, , 
\ee 
with effective mass $m = \pm \frac{\gamma}{\sqrt{2} v_F^2}$, or equivalently, recalling \eqref{eq:alpha}, $\alpha =\sqrt{2}$, 
a smaller value than for $AB$ bilayer graphene.
For the two insulating bands 
\be \label{eq:ABA_insulating_low_energy}
E_{ABA,2}^{(\pm)} \sim 2 m v_F^2 +  \frac{\hbar^2 |\kappa|^2}{2 m }  + O\left( \epsilon^4 \right) \, .  
\ee
This is the same gap, for given mass, found in the bilayer case. The energy is again in non-relativistic form and therefore we expect that non-relativistic fermions will appear. 
The full formula \eqref{eq:EABA} for the energy levels can be rewritten in terms of $m$ as 
\be \label{eq:EABA-2}
\hspace{-0.35cm} E_{ABA,i}^{(\pm)} = \pm \left[ 2 m^2 v_F^4 + \hbar^2 v_F^2 |\kappa|^2 + (-1)^i 2 m^2 v_F^2 \sqrt{1 + \frac{\hbar^2  |\kappa|^2}{m^2 v_F^2}} \right]^{\frac{1}{2}} \hspace{-0.25cm}, \quad i = 1, 2.  
\ee 
It is a different formula from \eqref{eq:energy_free1} but it has an identical low energy limit. 

The spinor solution of the equation $H\lambda = E \lambda$ is given by 
\ba \label{eq:sol_3layer_spinors1}
\lambda_1=1\, , \quad \lambda_2 = \frac{E}{\hv\bar{\kappa}} \, , \quad \lambda_3 = \sigma \alpha \frac{E}{\hv \bb{\kappa}} \, , \quad  
\lambda_4 =  \sigma \alpha \frac{\kappa}{\bb{\kappa}} \, ,  \quad \lambda_5 = 1 \, , \quad \lambda_6 = \frac{E}{\hv\bar{\kappa}} \, , 
\ea 
where $\sigma=\pm 1$ is a factor labelling different bands that appears in the consistency condition satisfied by the energy 
\be 
E^2 - |\hv \kappa |^2 - \sigma \alpha \gamma E = 0 \, . 
\ee

Now we analyse the spinor solutions close to the Fermi point for the metallic bands: 
we substitute the low energy limit \eqref{eq:ABA_metallic_low_energy} into the spinor solution, and retain the lowest order in the parameter $\epsilon$. The result is given by \eqref{eq:bilayer_metallic_lambda2}--\eqref{eq:bilayer_metallic_lambda4} plus 
\ba 
\lambda_5 &=& \frac{4}{\alpha^2} -1  \label{eq:trilayer_metallic_5} \, , \\ 
 \lambda_6 &=& \left( \frac{4}{\alpha^2} -1 \right) \frac{\hbar \kappa}{2 m v_F} + O (\epsilon^2) \, . \label{eq:trilayer_metallic_6}
\ea 
If we build $\chi_1$ and $\chi_2$ as in \eqref{LL_solution1_E1_tau+}, \eqref{LL_solution2_E1_tau+} then again these satisfy the L\'evy-Leblond equations. There is also another set of L\'evy-Leblond equations with spinors 
\ba 
\chi_1(t)  &=& 
 e^{- i \frac{\hbar |\kappa|^2 t}{2m}}  \left(    \lambda_3 , B \lambda_6 \right)^T \, , \\ 
\chi_2(t)  &=& 
 e^{- i  \frac{\hbar |\kappa|^2 t}{2m}}  \left(  B\lambda_5 ,  \lambda_4   \right)^T \, , 
\ea  
where $B$ is another arbitrary constant. As for the bilayer, at the Fermi point $\kappa = 0$ the $p_z$ orbitals of the atoms directly aligned $B1$, $A_2$, $B_3$ are turned off. So we see that the non-relativistic fermions for the metallic bands of three-layer graphene consist of states where interlayer hopping happens between all three layers, generalizing the physics occurring in bilayer graphene: interlayer hopping is a fundamental ingredient in order to generate a mass for the excitations. 
  
The same procedure, applied to the insulating bands  yields  the expansion \eqref{eq:bilayer_insulating_lambda2}--\eqref{eq:bilayer_insulating_lambda4}, plus the low energy limit 
\ba 
\lambda_5 &=& \frac{4}{\alpha^2} -1  \label{eq:trilayer_insulating_5} \, , \\ 
 \lambda_6 &=& \left( \frac{4}{\alpha^2} -1 \right) \frac{2 m v_F}{\hbar \bb{\kappa}} + O \left(\epsilon^2\right) \, . \label{eq:trilayer_insulating_6}
\ea 
This time at the Fermi point the contributions of the atoms $A_1$, $B_2$, $A_3$ are turned off. The L\'evy-Leblond spinors in this case are given by 
\ba 
\chi_1(t)  &=& 
 e^{- i \frac{\hbar |\kappa|^2 t}{2m}}  \left(  B\lambda_5 ,  \lambda_4   \right)^T  \, , \\ 
\chi_2(t)  &=& 
 e^{- i  \frac{\hbar |\kappa|^2 t}{2m}} \left(    \lambda_3 , B \lambda_6 \right)^T  \, .  
\ea  
 
We conclude this section analysing the bands that are linear in momentum. The solution close to the bottom of such Dirac bands is given by 
\ba 
&& \lambda_1 = 1 \, , \qquad \lambda_2 = \pm \sqrt{\frac{\kappa}{\bb{\kappa}}} \, , \\ 
&& \lambda_3 = 0 \, , \quad \lambda_4 =  0 \, , \\ 
&& \lambda_5 = -1  \, , \quad  \lambda_6 = \mp \sqrt{\frac{\kappa}{\bb{\kappa}}} \, . 
\ea 
Then the spinors 
\ba 
\psi^{(\pm)}_1(t) &=&  e^{ i v_F |\kappa| t} \left( \begin{array}{c} \lambda_1 \\  \lambda_2 \end{array} \right) \, , \\ 
\psi^{(\pm)}_2(t) &=&  e^{ i v_F |\kappa| t} \left( \begin{array}{c} \lambda_5 \\  \lambda_6 \end{array} \right) \, ,
\ea 
satisfy the Weyl equations for massless spinors 
\be 
\pa_t \psi^{(\pm)} = \pm v_F D \psi^{(\pm)} \, .  
\ee 
These solutions are localised on the layers $1$ and $3$, and they vanish in the intermediate layer $2$. On the layers, they describe massless modes moving at the speed of sound $v_F$. One can think of these solutions in this way: the electronic wavefunctions undergo totally destructive inteference in the middle layer, and the remaining electrons are confined to the top and bottom layers. Since interlayer hopping is forbidden in these localized states, then the excitations become massless again as for monolayer graphene. 
  
\subsection{Four-layer graphene\label{sec:four_layers}}  
The case of four-layer graphene with the ABAB stacking of the layers
is finally considered. Its schematic representation is reported
in figure \ref{fig:flat4}. 
The ABAB Hamiltonian $H_k^{(ABAB)}$ around the K point is given by 
\be \label{eq:ABAB_Hamiltonian} 
H_\kappa^{(ABAB)} = \left( 
\begin{array}{cccccccc} 
	0 & \hbar v_F \bb{\kappa} & 0 & 0 & 0 & 0 & 0 & 0 \\ 
	\hbar v_F \kappa & 0 & \gamma & 0 & 0 & 0 & 0 & 0\\ 
	0 & \gamma & 0 & \hbar v_F \bb{\kappa} & 0 & \gamma & 0 & 0\\ 
	0 & 0 & \hbar v_F \kappa & 0 & 0 & 0 & 0 & 0 \\ 
	0 & 0 & 0 & 0 & 0 & \hbar v_F \bb{\kappa} & 0 & 0\\ 
	0 & 0 & \gamma & 0 & \hbar v_F \kappa & 0  & \gamma & 0 \\ 
	0 & 0 & 0 & 0 & 0 & \gamma & 0 & \hbar v_F \bb{\kappa} \\ 
	0 & 0 & 0 & 0 & 0 & 0 & \hbar v_F \kappa & 0 \\ 
\end{array} 
\right) \, . 
\ee 
This is the same as for the three-layer case, with the addition of atoms $B_3$, $A_4$ directly aligned. We plain to repeat the analysis done in the previous section: first finding the shape of the bands, then looking for special types of excitations. In this case we will find only L\'evy-Leblond spinors, and no massless excitations. 
 
Analysing the eigenvalues of $H_k^{(ABAB)}$ we find 8 bands, expressed in terms of two masses: 
\ba 
m_1 &=& \frac{\sqrt{5} -1}{4} m_0 \sim 0.309 \, m_0 \, , \\ 
m_2 &=& \frac{\sqrt{5} +1}{4} m_0  \sim 0.809 \,  m_0 \, . 
\ea 
In our notation $\alpha_1 = \frac{4}{\sqrt{5}-1} \sim 3.236$, $\alpha_2 = \frac{4}{\sqrt{5}+1} \sim 1.236$. 
For each mass $m_i$ there are two metallic and two insulating bands. For $m_1$: 
\be \label{eq:energy_free1_4layers}
E_{i,m_1}^{(\pm)} = \pm \left[ (-1)^{i} m_1 v_F^2 + \sqrt{ m_1^2 v_F^4 + |\hv\kappa|^2} \right] \, , i = 1,2 , 
\ee  
And for $m_2$ 
\be \label{eq:energy_free2_4layers}
E_{i,m_2}^{(\pm)} = \pm \left[ (-1)^{i} m_2 v_F^2 + \sqrt{ m_2^2 v_F^4 + |\hv\kappa|^2} \right] \, , i = 1,2 , 
\ee 
The components of the energy eigenspinors are given by 
\ba \label{eq:sol_4layer_spinors1}
\lambda_1 &=&1\, ,\quad\lambda_2 = \frac{E}{\hv\bar{\kappa}} \, , \quad \lambda_3 = \sigma \frac{2}{\alpha_i}  \frac{E}{\hv \bb{\kappa}} \, , \quad  
\lambda_4 =  \sigma \frac{2}{\alpha_i} \frac{\kappa}{\bb{\kappa}} \, ,  \quad 
\lambda_5 = \left(\frac{4}{\alpha_i^2}-1\right) \, ,  \\  \lambda_6 &=& \left(\frac{4}{\alpha_i^2}-1\right) \frac{E}{\hv\bar{\kappa}} \, , \quad 
\lambda_7 = \sigma \frac{2}{\alpha_i}  \left(\frac{4}{\alpha_i^2}-1\right) \frac{E}{\hv\bar{\kappa}} \, , \quad \lambda_8 = \sigma \frac{2}{\alpha_i}  \left(\frac{4}{\alpha_i^2}-1\right) \frac{\kappa}{\bar{\kappa}} \, , 
\ea 
where $\sigma=\pm 1$ is a factor labelling different bands that appears in the consistency condition satisfied by the energy 
\be 
E^2 - |\hv \kappa |^2 - \sigma \frac{2}{\alpha_i} \gamma E = 0 \, . 
\ee 
At the bottom of the metallic bands $E_{1,m_i}^{(\pm)}$ band they take the form \eqref{eq:bilayer_metallic_lambda2}-\eqref{eq:bilayer_metallic_lambda4},  \eqref{eq:trilayer_metallic_5}-\eqref{eq:trilayer_metallic_6} plus 
\ba 
&& \hspace{-1cm} \lambda_7 = \mp \frac{2}{\alpha_i} \left(\frac{4}{\alpha_i^2} - 2 \right) \frac{\hbar \kappa}{2m_i v_F}+O\left(\epsilon^2\right) \, , \\ 
&& \hspace{-1cm} \lambda_8 = \mp \frac{2}{\alpha_i} \left(\frac{4}{\alpha_i^2} - 2 \right) \frac{\kappa}{\bb{\kappa}} \, . 
\ea 
Similarly, at the bottom of the insulating bands $E_{2,m_i}^{(\pm)}$ the spinor takes the form \eqref{eq:bilayer_insulating_lambda2}-\eqref{eq:bilayer_insulating_lambda4}, \eqref{eq:trilayer_insulating_5}-\eqref{eq:trilayer_insulating_6} plus 
\ba 
&& \hspace{-1cm} \lambda_7 = \pm \frac{2}{\alpha_i} \left(\frac{4}{\alpha_i^2} - 2 \right) \frac{2m_i v_F}{\hbar \bb{\kappa}}+O\left(\epsilon^2\right)  \, , \\ 
&& \hspace{-1cm} \lambda_8 = \pm \frac{2}{\alpha_i} \left(\frac{4}{\alpha_i^2} - 2 \right) \frac{k}{\bb{\kappa}} \, . 
\ea 
Therefore we see repeating the same pattern that we found for the bilayer and three-layer cases: at the Galilei point for the metallic bands the component related to the $A_4$ atom is turned off, together with those of all the atoms that are directly aligned, and at the Galilei point for the insulating bands conversely the $B_4$ component is zero. 
 
The new L\'evy-Leblond spinors for the metallic bands are 
\ba 
\chi_1(t)  &=& 
 e^{- i \frac{\hbar |\kappa|^2 t}{2m_i}}  \left(  C\lambda_7 ,  \lambda_6   \right)^T  \, , \\ 
\chi_2(t)  &=& 
 e^{- i  \frac{\hbar |\kappa|^2 t}{2m_i}} \left(    \lambda_5 , C \lambda_8 \right)^T  \, , 
\ea   
while for the insulating bands 
\ba 
\chi_1(t)  &=& 
 e^{- i \frac{\hbar |\kappa|^2 t}{2m_i}}  \left(  \lambda_5 , C \lambda_8 \right)^T   \, , \\ 
\chi_2(t)  &=& 
 e^{- i  \frac{\hbar |\kappa|^2 t}{2m_i}} \left(  C\lambda_7 ,  \lambda_6   \right)^T  \, ,   
\ea 
where $C$ is another arbitrary constant. We see here that with four layers, no configurations with totally destructive interference in the middle layers appear. All states present interlayer hopping and are massive.

\section{The curved case\label{sec:3}}  
\subsection{The L\'evy-Leblond equation} 
\begin{figure}%
         \centering
         \subfloat[$\lambda_1=\lambda_2=0$]{\label{fig:superpos-a}\includegraphics[width=0.45\textwidth]{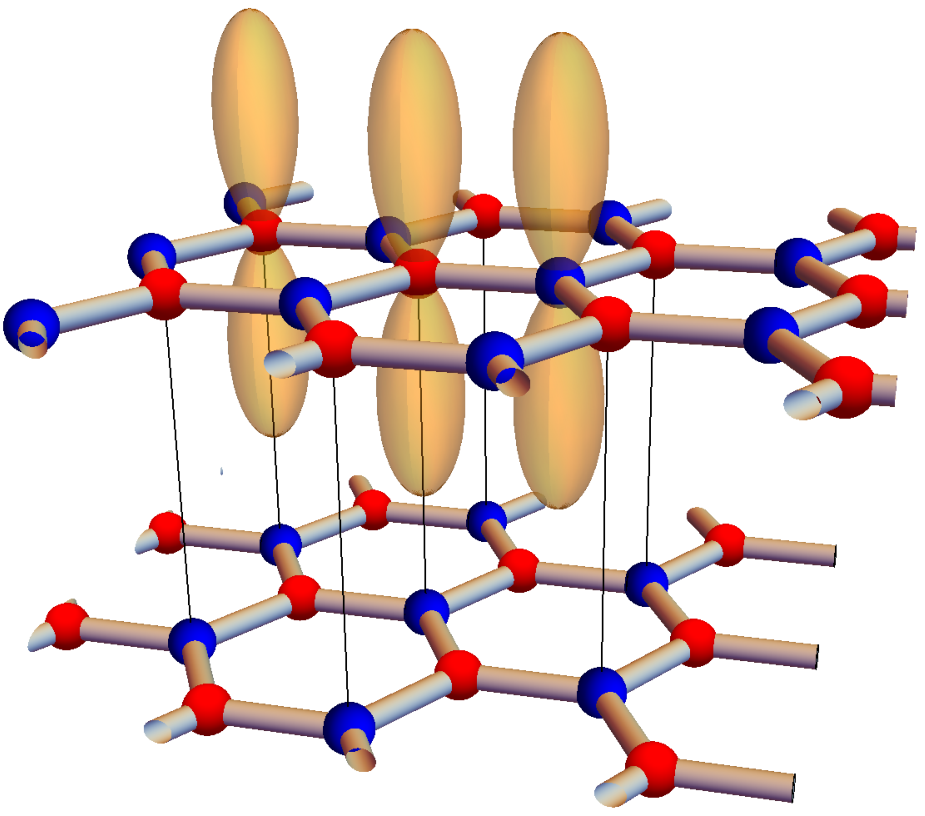}}\qquad
         \subfloat[$\lambda_3=\lambda_4=0$]{\label{fig:superpos-b}
         \includegraphics[width=0.45\textwidth]{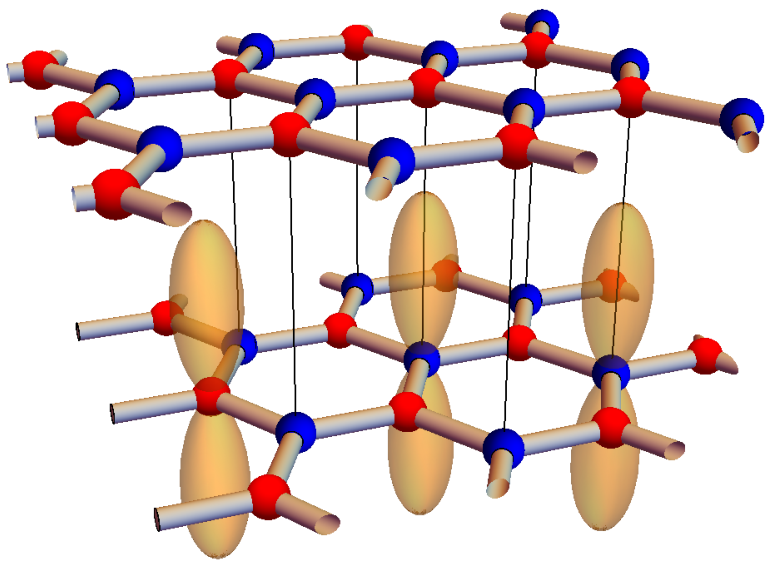}} 
         \caption{States with definite pseudospin in bilayer graphene close to the Galilei points. From eqs.\eqref{LL_solution1_E1_tau+}, \eqref{LL_solution2_E1_tau+}, \eqref{LL_solution1_E2_tau+}, \eqref{LL_solution2_E2_tau+} it can be seen that for both the metallic and insulating bands the two eigenstates of pseudospin have either $\lambda_1 = 0 = \lambda_2$, or $\lambda_3 = 0 = \lambda_4$. These states correspond to excitations concentrated in only one of the two planes. }
         \label{fig:superpos}
       \end{figure}

The L\'evy-Leblond equations \eqref{eq:LevyLeblond1}, \eqref{eq:LevyLeblond2} are coupled, first order equations. L\'evy-Leblond in his seminal paper \cite{LevyLeblond1967} showed that these are obtained from the theory of representations of the Galilei group applied to particles of spin $\frac{1}{2}$. In general, the L\'evy-Leblond equations can be solved in the following way: solving for $\chi_1$ in \eqref{eq:LevyLeblond2} and plugging it into \eqref{eq:LevyLeblond1} we obtain the second order, uncoupled equation
\be \label{eq:Schrodinger_full} 
i \hbar \, \pa_t \chi_2 =  - \frac{\hbar^2}{2m} \nabla^2  \chi_2 \, , 
\ee 
which in fact is the Pauli equation in the special case of the external magnetic field set to zero, see \cite{Curvatronics2017} for the case of a non-zero  field. This equation admits the  independent solutions 
\be \label{eq:chi2_solution_flat}
\chi_{2,\uparrow} = e^{- \frac{i \hbar |\kappa|^2 t}{2m} } \left( \begin{array}{c} 1 \\ 0 \end{array} \right)  \, , \qquad 
\chi_{2,\downarrow} = e^{- \frac{i \hbar |\kappa|^2 t}{2m} } \left( \begin{array}{c} 0 \\ 1 \end{array} \right) 
\ee 
which correspond to different values of $2D$ (pseudo)spin. Then using \eqref{eq:LevyLeblond2} $\chi_1$ is obtained by differentiation of $\chi_2$ as 
\be 
\hspace{-0.2cm} \chi_{1,\downarrow} = \frac{\hbar}{2 m v_F} e^{- \frac{i \hbar |\kappa|^2 t}{2m} } \left( \begin{array}{c} 0 \\ \kappa \end{array} \right)  \, , \quad 
\chi_{1,\uparrow} = \frac{\hbar}{2 m v_F} e^{- \frac{i \hbar |\kappa|^2 t}{2m} } \left( \begin{array}{c} \bb{\kappa} \\ 0 \end{array} \right) 
\ee
By linearity one can then construct the full solution: 
\ba 
\chi_1 &=& a \chi_{1,\downarrow} + b \chi_{1,\uparrow} \, , \\ 
\chi_2 &=& a \chi_{2,\uparrow} + b \chi_{2,\downarrow} \, . 
\ea 
Let us compare this general formula with, for example, the L\'evy-Leblond spinor obtained for the metallic bands $E_1^{(\pm)}$  in the case of bilayer graphene: 
\ba 
 \chi_1(t)  &=& 
 e^{- i \frac{\hbar |\kappa|^2 t}{2m}}  \left( \mp  \frac{2}{\alpha} \frac{\hbar \kappa}{2mv_F} ,  \frac{\hbar \kappa}{2mv_F} \right)^T \, , \\ 
\chi_2(t)  &=& 
 e^{- i  \frac{\hbar |\kappa|^2 t}{2m}}  \left(  1 , \mp  \frac{2}{\alpha} \frac{\kappa}{\bb{\kappa}}   \right)^T \, .  
\ea 
This can be obtained from the general formula setting $a=1$, $b = \mp  \frac{2}{\alpha}\frac{\kappa}{\bb{\kappa}}$. So the L\'evy-Leblond spinor that occurs for a definite value of the energy is given by a linear superposition of spinors with definite values of the pseudospin. Conversely, a linear superposition of L\'evy-Leblond spinors from the bands $E_1^{(+)}$ and $E_1^{(-)}$ corresponds to states of definite value of the pseudospin. Similar formulae hold for L\'evy-Leblond spinors related to the insulating bands, as well as for the three- and four-layer case. States with definite pseudospin in the case of bilayer graphene close to the Galilei points are schematically shown in Figure \ref{fig:superpos}. 

We now recall the relationship with the massless Dirac equation in $4$ dimensions. It has been shown in \cite{Curvatronics2017} that given a solution $\left( \chi_1, \chi_2 \right)$ of the L\'evy-Leblond equations it is possible to construct a massless Dirac spinor in an effective 4--dimensional Minkowski spacetime 
\be \label{eq:Minkowski}
g_{\mu\nu} dx^\mu dx^\nu = dx^2 + dy^2 + 2 du dv \, ,    
\ee 
where $u$, $v$ are conjugate null coordinates. The massless 4D spinor is given by 
\be \label{eq:spinor_4d_decomposition}
\h{\Psi}(u,v,x,y) = e^{i \frac{m v_F}{\hbar} v}   \left( \begin{array}{c} \chi_1(u,x,y) \\ \chi_2(u,x,y) \end{array} \right) \, . 
\ee 
In terms of degrees of freedom, the massless Dirac equation in 4D has two degrees of freedom, corresponding to two different allowed chiralities. In turn, the L\'evy-Leblond equations admit two independent solutions, as just seen. This fact is well known in the literature \cite{Duval1985,Duval1995_1,Duval1995_2,Cariglia2012}, and is related to the concept of the Eisenhart-Duval lift of non-relativistic mechanics \cite{Eisenhart1928,DBKP,DGH91}. The relationship can be clearly seen in terms of the Gamma matrices adapted to the 4D lift: 
\ba  
\Gamma^+ &=&    \left( \begin{array}{cc} 0 & \mathbb{I} \\ 0 & 0 \end{array} \right)  , \,  \qquad \Gamma^- =   2 \left( \begin{array}{cc} 0 & 0 \\ \mathbb{I} & 0 \end{array} \right)  , \, \qquad  
\Gamma^i =   \left( \begin{array}{cc} \sigma_i & 0 \\ 0 & - \sigma_i \end{array} \right) \, . \nn 
\ea  
In terms of these the chirality matrix is calculated as 
\be \label{eq:Gamma_star}
\Gamma^* = i \, \Gamma^0 \, \Gamma^1 \, \Gamma^2 \, \Gamma^3 = \left( \begin{array}{cc} - \sigma_3 & 0 \\ 0 & \sigma_3 \end{array} \right) \, , 
\ee 
showing that the spinor $\h{\Psi}$ built from the combination $a=1$, $b=0$ has a definite chirality, which by convention we can call right, and the one built using $a=0$, $b=1$ has opposite chirality, say left. That chirality in 4D can be related to pseudospin in 2D can be understood in the following way. First, we notice that in 4D we can introduce a timelike variable $t$ and a spacelike variable $z$ according to $2dudv = -dt^2 + dz^2$. Then, from eq.\eqref{eq:Gamma_star} chirality is decomposed into a boost in the $z$ direction, times a rotation around the $z$ axis. Spinors of the form \eqref{eq:spinor_4d_decomposition} with either $\chi_1 = 0$ or $\chi_2 = 0$ are eigenspinors of boosts along the $z$ direction: this  is related to the fact that the metric \eqref{eq:Minkowski} is written in null coordinates $u, v$ and that the spinors are adapted to the coordinates. So for these spinors 4D chirality and 2D isospin are directly related.

\subsection{The role of curvature} 
We now consider curved few layers of graphene. Our chief example will be the bilayer case, however similar considerations apply to the case of three or four layers.  
When the radius of curvature is small enough we consider a Hamiltonian that satisfies the following two requirements. First, the Hamiltonian should be self-adjoint, so that the energy eigenvalues are real. Second, it should be a covariant generalization of \eqref{eq:H}, which reduces to \eqref{eq:H} when the metric is flat. In order to do so, we replace the momentum $k$ with $-i \nabla$, the 2D spinorial covariant derivative that includes the spin connection of the curved metric $g_{ij} dq^i dq^j$, 
with $q^i = \{x,y\}$.

\begin{figure}%
         \centering
         \subfloat[spherical]{\label{fig:elliptic}\includegraphics[width=0.45\textwidth]{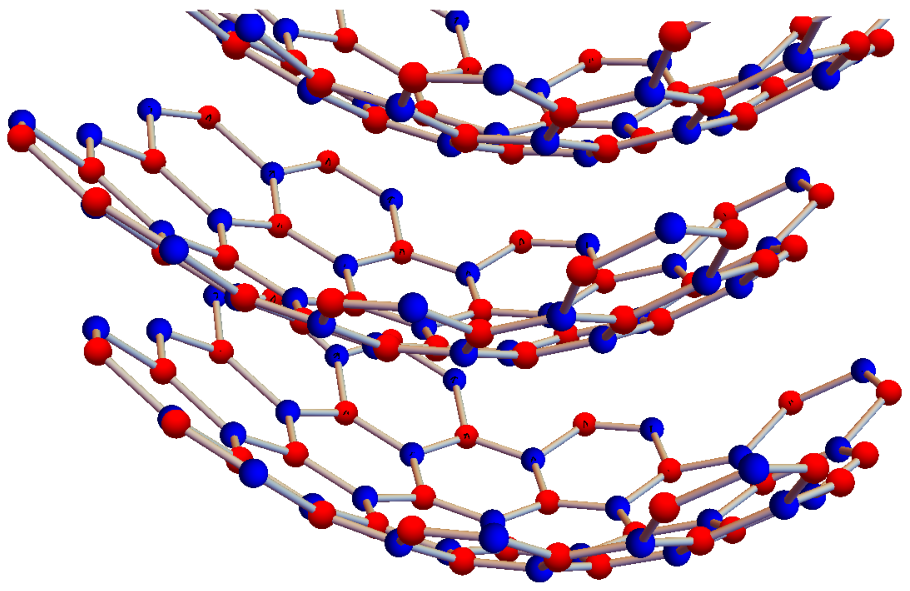}}\qquad
         \subfloat[hyperbolic]{\label{fig:hyperbolic}
         \includegraphics[width=0.45\textwidth]{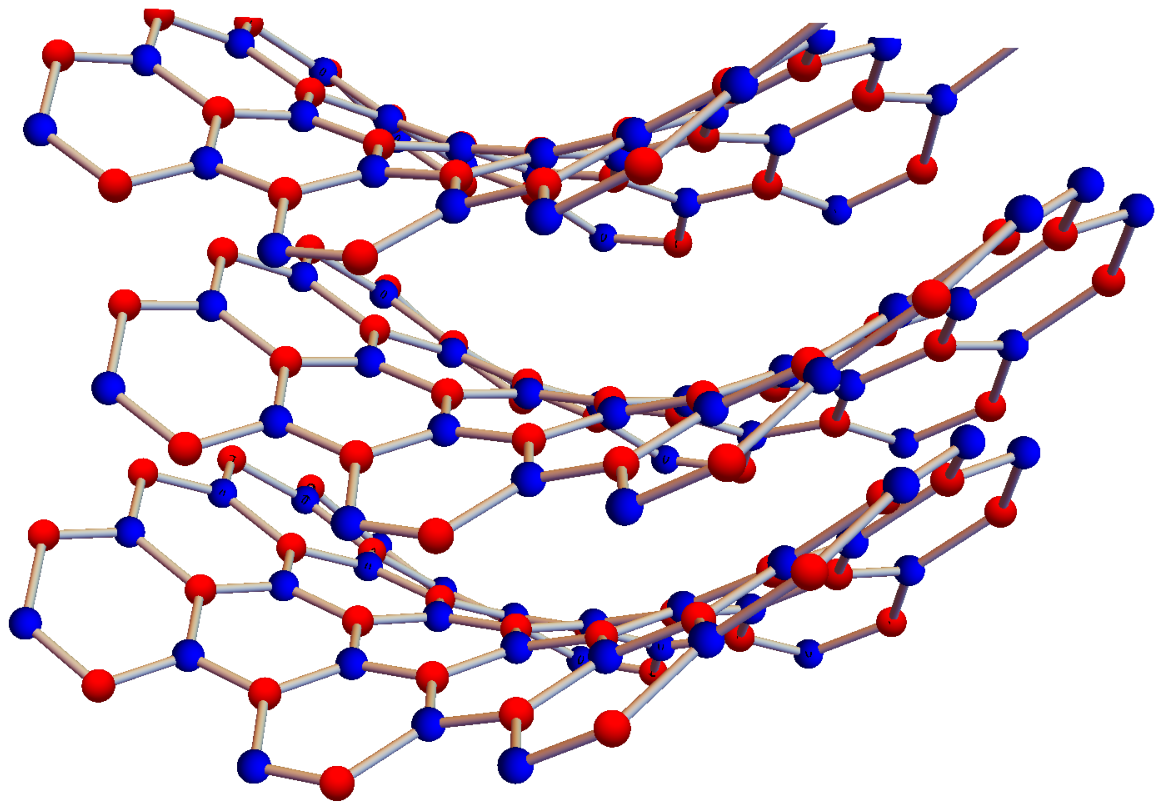}} 
         \caption{Configurations with curvature for the three-layer graphene system exploitable for curvatronics, to tune the electronic properties of few-layer graphene by geometrical effects. (a) Spherical (positive) curvature. (b) Hyperbolic (negative) curvature.}
         \label{fig:3curved}
  \end{figure}
In sec. \ref{sec:two_layers} we have seen how the L\'evy-Leblond spinors appeared reorganizing the spinor components as per \eqref{LL_solution1_E1_tau+}. Motivated by this insight  we  make the following change of basis 
\ba 
|1\rangle &=& |A2\rangle \, \label{eq:change1} \\ 
|2\rangle &=& |B1\rangle \,  \\ 
|3\rangle &=& |A1\rangle  \,   \\ 
|4\rangle &=& |B2\rangle  \label{eq:change4}  \, , 
\ea 
represented by the following linear function $R: \{A1,B1,A2,B2\} \mapsto \{1,2,3,4\}$ 
\be R = \left( \begin{array}{cccc}  
0 & 0 & 1 & 0 \\ 
0 & 1 & 0 & 0 \\ 
1 & 0 & 0 & 0 \\ 
0 & 0 &0 & 1 
\end{array}\right). 
\ee 
The Hamiltonian changes accordingly into  
\ba \label{eq:H_transformation_rule}
H^\prime_{\kappa} &=& R H_{\kappa} R^{-1} 
\ea 
or 
\ba \label{eq:newH}
H^\prime_{\kappa} &=& 
\left( \begin{array}{c|c} 
\begin{array}{cc} 0 & \gamma   \\ \gamma & 0 \end{array} 
& - i \hv D \\ \hline 
 - i \hv D & 
\begin{array}{cc} 0 & 0  \\ 0 & 0 \end{array} 
\end{array} \right) \, ,   
\ea 
which is (formally) self-adjoint since $D^\dagger = - D$. The electronic wavefunction is 
\ba 
|\Psi\rangle &=& \lambda_1 |A1\rangle + \lambda_2 |B1\rangle + \lambda_3 |A2\rangle + \lambda_4 |B2 \rangle \nn \\ 
&=& \lambda_3 |1\rangle + \lambda_2 |2\rangle + \lambda_1 |3\rangle + \lambda_4 |4\rangle \, , 
\ea 
and similarly to what we have done in the flat case we regroup the components into 
\ba 
|\chi_1\rangle = \left( \begin{array}{c} \lambda_3 \\ \lambda_2 \end{array} \right) \, , 
\ea 
\ba 
|\chi_2\rangle = \left( \begin{array}{c} \lambda_1 \\ \lambda_4 \end{array} \right) \, . 
\ea 
The eigenvalue equation for $H^\prime_{\kappa}$ becomes 
\ba 
E \chi_1  &=& \left(  \begin{array}{cc}  0 & \gamma  \\  \gamma & 0 \end{array} \right) \chi_1 - i \hv D \chi_2  \\ 
E \chi_2 &=&  - i \hv D \chi_1 \, . 
\ea 
We can decouple the equations by obtaining $\chi_2$ from the second, and inserting it into the first.
 
An important case is that of metallic bands, for which the $E\chi_1$ term in the top equation is negligible with respect to the other two: in this limit 
\ba \label{eq:correct1}
&& 0  = \left(  \begin{array}{cc}  0 & \gamma  \\  \gamma & 0 \end{array} \right) \chi_1 - i \hv D \chi_2  \\ 
&& E \chi_2 = - i \hv D\chi_1 \, , \qquad (\mbox{metallic bands}).  \label{eq:correct2}
\ea 

Recalling that $\gamma = 2 |m| v_F^2$, these equations are similar to the stationary L\'evy-Leblond equations, with the difference that the term $-2mv_F^2$ is replaced by the matrix \begin{small}$\left(  \begin{array}{cc}  0 & \gamma  \\  \gamma & 0 \end{array} \right)$\end{small}. To understand the relationship with what we found in the flat case, we isolate $\chi_2$ in \eqref{eq:correct2}, plug it into \eqref{eq:correct1} and obtain 
\be \label{eq:chi1_quadratic}
E\chi_1 = \frac{\hbar^2 v_F^2}{\gamma} \left(  \begin{array}{cc}  0 & 1  \\  1 & 0 \end{array} \right) D^2 \chi_1 \, . 
\ee 

This is the eigenvalue equation for the energy in the curved case. In flat space and for metallic bands we found $E=\pm \frac{\hbar^2 v_F^2 |\kappa|^2}{\gamma}$ so that  \eqref{eq:chi1_quadratic} became  
\be 
\chi_1 = \left(  \begin{array}{cc}  0 & \mp1  \\  \mp1 & 0 \end{array} \right) \chi_1 \, , 
\ee 
and in fact our solution was  
\ba \label{eq:chi1_flat_solutions}
\chi_1 = \frac{\hbar}{2m v_F \kappa} \left( \begin{array}{c} \mp1 \\ 1 \end{array} \right) \, . 
\ea 
Notice that in this case $\chi_1$ is an eigenspinor of the matrix $\sigma_1$ with eigenvalues $\mp 1$, so that we are justified in exchanging the matrix \begin{small}$\left(  \begin{array}{cc}  0 & \gamma  \\  \gamma & 0 \end{array} \right)$\end{small} with $-2mv_F^2$. However, in general the operator on the right hand side of \eqref{eq:chi1_quadratic} does not commute with $\sigma_1$, and as a consequence eqs.(\ref{eq:correct1}-\ref{eq:correct2}) are a variation of the L\'evy-Leblond equations. 
 
Our theory predicts that there will appear a gap in the spectrum if there are no solutions of \eqref{eq:chi1_quadratic} for the eigenvalue $E=0$. In this particular case the analysis reduces to finding  zero modes of $D$ , and it is a known result that if the surface has positive curvature then the spectrum of $D$ does not include the eigenvalue zero. 
 
Eq.\eqref{eq:chi1_quadratic} is in general non-trivial to study, and to gain a better understanding we focus our attention on the axisymmetric case. That is, we consider a metric of the type 
\be 
g_{ij} dq^i dq^j = dr^2 + C(r)^2 d\varphi^2 \, . 
\ee 
For example for $C(r)^2 = \sin(r)^2$ the geometry is that of a sphere, with constant positive curvature, and for $C(r)^2 = r^2 + a^2$, with $a$ a constant, the geometry describes two flat ends joined by a tube, and has negative, non-constant, curvature. In the generic axisymmetric case the Dirac operator is given by 
\ba 
D &=& \sigma_1 \left( \pa_r + \frac{C^\prime}{2 C} \right) + \sigma_2 \frac{\pa_\varphi}{C} \nn \\ 
&=& \left( \begin{array}{c|c} 
0 & \mathcal{O} \\ 
\hline - \mathcal{O}^\dagger & 0
\end{array} \right)
\ea 
where 
\ba 
\mathcal{O} &=& \pa_r + \frac{C^\prime}{2C} + \frac{-i\pa_\varphi}{C} \nn \\ 
\mathcal{O}^\dagger &=& -\pa_r - \frac{C^\prime}{2C} + \frac{-i\pa_\varphi}{C} \, , 
\ea
so that  
\be 
D^2 = -
\left( \begin{array}{c|c} 
\mathcal{O} \mathcal{O}^\dagger & 0 \\ 
\hline 0 &  \mathcal{O}^\dagger \mathcal{O} 
\end{array} \right) \, . 
\ee 

We will also assume that the operator acts on a spinor  of the form $\chi_1 = e^{i l \varphi} \left(\begin{array}{c} a(r) \\ b(r) \end{array}\right)$, $l = \pm \frac{1}{2}, \pm \frac{3}{2}, \dots$. From our calculation then eq.\eqref{eq:chi1_quadratic} reduces to the pair of equations 

\ba 
\frac{\gamma E}{\hbar^2 v_F^2}  a &=& - B b \, ,  \label{eq:gammaE_1} \\ 
\frac{\gamma E}{\hbar^2 v_F^2}  b &=& - A a \, , \label{eq:gammaE_2} \, 
\ea 
where we defined the second order, self-adjoint, positive operators $A = \mathcal{O} \mathcal{O}^\dagger$, $B = \mathcal{O}^\dagger \mathcal{O}$. These equations can be decoupled,  giving  
\ba 
\frac{\gamma^2 E^2}{\hbar^4 v_F^4} a &=& B A \, a \, , \label{eq:eigenvalue_a} \\ 
\frac{\gamma^2 E^2}{\hbar^4 v_F^4}  b &=& A B \, b  \label{eq:eigenvalue_b} \, ,  
\ea 
or in other words on the space of $\chi_1$ spinors the operator $\gamma^2 E^2$ is represented by 
\be \label{eq:energy_squared}
\frac{\gamma^2 E^2}{\hbar^4 v_F^4} = \left( \begin{array}{c|c} 
BA & 0 \\ 
\hline 0 &  AB 
\end{array} \right) \, . 
\ee 
To summarize, we obtained a fourth order equation for the square of the energy eigenvalues. Our equation is consistent since it is a known fact that for positive operators $A$ and $B$ the spectrum of $AB$ is the same of that of $BA$, and it is positive \cite{Hladnik1988}.  Being fourth order, our equations will be challenging to solve analytically in general: the solution will require a careful but feasible numerical approach, which we postpone to future works. There are however some properties that are easy to discuss. First, since they are quadratic in $E$, then the spectrum will be symmetric under $E \rightarrow -E$. Second, as the operator in \eqref{eq:energy_squared} commutes with the matrix $\sigma_3$, then we can see that if $a$, $b$ are eigenfunctions related to the eigenvalue $E$, then $a$, $-b$ are eigenfunctions related to the eigenvalue $-E$. Thus, in analogy to the flat case, we are able to construct checkered solutions  from linear combinations of solutions with opposite energy, and viceversa. 
 
However, in general it will be the case that $a \neq b$, since $A \neq B$. Therefore the solutions cannot  be related to L\'evy-Leblond fermions. There is however a regime where we reasonably expect that $A  \sim B$. One can calculate $\left[\mathcal{O},\mathcal{O}^\dagger \right] = - 2l \frac{C^\prime}{C^2}$. Suppose the function $\left| 2 \frac{C^\prime}{C^2}\right|$ has a maximum $\mu$. We can substitute $B$ in \eqref{eq:gammaE_1} for $A + \left[\mathcal{O}^\dagger,\mathcal{O} \right]$ and, in those cases where 
\be 
\frac{\gamma E}{\hbar^2 v_F^2} >> l \mu 
\ee 
we can ignore the extra term and reduce the eigenvalue equations to 
\ba 
\frac{\gamma E}{\hbar^2 v_F^2}  a &=& - A b \, , \label{eq:coupled_1}  \\ 
\frac{\gamma E}{\hbar^2 v_F^2}  b &=& - A a  \label{eq:coupled_2} \, .  
\ea
These admit two types of solutions: solutions with $b = a$ give 
\be 
\frac{\gamma E}{\hbar^2 v_F^2}  a = - A a \, , 
\ee 
i.e. $a$ is an eigenfunction of $A$ and the energy is negative, or $b = - a$ which gives 
\be 
\frac{\gamma E}{\hbar^2 v_F^2}  a =  A a \, , 
\ee
i.e. $a$ is an eigenfunction of $A$ and the energy is positive. In this approximation of low angular momentum to energy ratio the problem simplifies considerably, as one needs to solve 2nd order differential equations. Summarising our results, we expect that deforming the few--layer graphene sheets provides an efficient way to increase or reduce an energy band gap between conduction and valence bands, since few layers graphene has several degrees of freedom with respect to which deformations can be introduced: both along the sheets and perpendicular to it. As an illustrative example, possible positive or negative curvatures
of the three-layer graphene are depicted in Figure \ref{fig:3curved}.



\section{Conclusions} 
In this work we have extended the geometrical approach and the results of our previous work on curvatronics with bilayer graphene of Ref.\cite{Curvatronics2017} to the case of three and four layers of AB stacked graphene, and we have refined the role of isospin states in the case of the curved bilayer. We found that for flat layers L\'evy-Leblond fermions are still present at the bottom of the parabolic bands, which we call Galilei points of the Brillouin zone, and we gave a geometrical interpretation of the solutions in terms of the $p_z$ orbitals of graphene associated to them, on the basis of their non-trivial spatial overlap. We have discussed in detail the effective electronic masses present for three and four layers, comparing them with the bilayer graphene case, as well as the explicit form of the solutions. We analysed the L\'evy-Leblond equation and showed how energy eigenstates are given by the superpositions of checkered states. Using the relationship between the 2D L\'evy-Leblond equation and the 4D massless Dirac equation, we were able to relate 2D pseudospin with chirality in 4D. Lastly, we discussed the fact that curving few layers graphene provides a tool to tune the energy band gap, which has implications for tunable electronics based on curvature, or curvatronics with few-layer graphene systems. For instance, a graphene--based device with alternating regions of positive or negative curvature, inducing insulating or conductive states, could be exploited to produce non--volatile memories. It would be interesting for future research to study in detail the consequences of having curved few layers, as well as generalize our results to a higher number of layers, and to study  the fermionic excitations away from the Galilei points.  Another avenue of possible future research is studying the physics of pseudospin in settings with non-trivial curvature or topology, which might lead to novel effects. 
 
Finally, very recently the class of Weyl metamaterials has been introduced in
Ref.\cite{Weststrom}. In these systems, chiral Weyl fermions
are moving in an artificial 3D curved-space geometry, with applications to
tunable novel electronic devices through curvature engineering and with
connections to cosmological theories. Therefore, the new field of
quantum physics in curved spaces is likely to become of great and practical interest
in the near future.

\vspace{0.2cm}

\textbf{Acknowledgments} -- 
M. Cariglia is funded by CNPq under project 303923/2015-6, and by a \textit{Pesquisador Mineiro} project n. PPM-00630-17. The authors acknowledge the collaboration within the MultiSuper International Network (http://www.multisuper.org) for exchange of ideas and suggestions.

\vspace*{-1ex}


\end{document}